\documentclass[prl,twocolumn,superscriptaddress,floatfix,nopacs]{revtex4}
\usepackage{graphicx,amsfonts,amssymb,amsmath,hyperref,enumerate,color}
\usepackage{bbold}
\usepackage{xspace}

\definecolor{augustine}{rgb}{1,0,0}
   \newcommand{\ak}{\textcolor{black}}  
   \newcommand{\je}{\textcolor{black}}
   \newcommand{\cb}{\textcolor{black}}
\newif\ifhyper
\hypertrue

\ifhyper
\hypersetup{
   citecolor = {green},
   colorlinks = {true}, 
   urlcolor = {blue} 
}
\fi

\newcommand{\beq}{\begin{equation}}
\newcommand{\eeq}{\end{equation}}
\newcommand{\beqa}{\begin{eqnarray}}
\newcommand{\eeqa}{\end{eqnarray}}

\newcommand{\blue}[1]

\def\Longarrow{\protect\@lra}
\def\@lra{\relbar\joinrel\relbar\joinrel\relbar\joinrel%
          \relbar\joinrel\rightarrow}

\usepackage{times}
\begin{document}

\title{Tensor network investigation of the double layer Kagome compound Ca$_{10}$Cr$_7$O$_{28}$} 

\author{Augustine Kshetrimayum}
\affiliation{Dahlem Center for Complex Quantum Systems, Physics Department, Freie Universit\"{a}t Berlin, 14195 Berlin, Germany}

\author{Christian Balz}
\affiliation{Neutron Scattering Division, Oak Ridge National Laboratory, Oak Ridge, TN 37831, USA}
\affiliation{ISIS Neutron and Muon Source, STFC Rutherford Appleton Laboratory, Didcot OX11 0QX, UK}

\author{Bella Lake}
\affiliation{Helmholtz-Zentrum Berlin f{\"u}r Materialien und Energie, 14109 Berlin, Germany}
\affiliation{Institut f{\"u}r Festk{\"o}rperphysik, Technische Universit{\"a}t Berlin, 10623 Berlin, Germany}

\author{Jens Eisert}
\affiliation{Dahlem Center for Complex Quantum Systems, Physics Department, Freie Universit\"{a}t Berlin, 14195 Berlin, Germany}
\affiliation{Department of Mathematics and Computer Science, Freie Universit\"{a}t Berlin, 14195 Berlin, Germany}
\affiliation{Helmholtz-Zentrum Berlin f{\"u}r Materialien und Energie, 14109 Berlin, Germany}

\begin{abstract}
Quantum spin liquids are exotic quantum phases of matter that do not order even at zero temperature. While there are several toy models and simple Hamiltonians that could host a quantum spin liquid as their ground state, it is very rare to find actual, realistic materials that exhibits their properties. At the same time, the classical simulation of such instances of strongly correlated systems is intricate and reliable methods are scarce. In this work, we investigate the quantum magnet Ca$_{10}$Cr$_7$O$_{28}$ that has recently been discovered to exhibit properties of a quantum spin liquid in inelastic neutron scattering experiments. This compound has a distorted bilayer Kagome lattice crystal structure consisting of Cr$^{5+}$ ions with spin-$1/2$ moments. Coincidentally, the lattice structure renders a tensor network algorithm in 2D applicable that can be seen as a new variant of a projected entangled simplex state algorithm in the thermodynamic limit. In this first numerical investigation of this material that takes into account genuine quantum correlations, good agreement with the experimental findings is found. We argue that this is one of the very first studies of physical materials in the laboratory with tensor network methods, contributing to uplifting tensor networks from conceptual tools to methods to describe real two-dimensional quantum materials.
\end{abstract}

\date{\today}

\maketitle

Since Anderson's proposal of the resonating valence bond state as one of the possible mechanisms of high $T_c$ superconductivity \cite{AndersonRVB}, the study of 
quantum spin liquids has attracted intense attention, both from theorists as 
well as experimentalists \cite{Balentsspinliquid,kanodaqsl,normandqsl,Coldeaqsl,WenKagomeqsl,j1j2pepsqsl}. Such phases of matter, besides falling beyond the paradigm of Landau symmetry breaking theory, 
exhibit intriguing physical properties such as fractional statistics of their excitations or a degeneracy of the ground state manifold that depends on the topology of the problem,
and feature no local order parameter. They are seen as key building blocks in topological quantum computing \cite{kitaevqsl} where it is the absence of
local order parameters that protects non-locally encoded quantum information against
local noise and errors. While several theoretical models have been proposed that could host quantum spin 
liquids as ground states \cite{j1j2pepsqsl,Andersonjij2,WhiteKagomeqsl}, finding actual materials natively 
exhibiting such properties which can be studied in a laboratory seem 
rare. This is mostly because realizing a quantum spin liquid 
requires complex frustration mechanisms which often need to be fine tuned. For this reason,
it is imperative to identify numerical simulation methods that can constructively guide such 
efforts. 

Meanwhile, the rapid development of novel, advanced theoretical and numerical techniques for studying 
complex quantum many-body systems continues to provide strong impetus to understanding problems such as the one described above. While there are several different techniques available for this purpose, many of them do not explicitly take 
into account the most important ingredient in strongly correlated models in general and quantum 
spin liquids in particular: These are genuine \emph{quantum correlations} in the 
ground state \cite{MF,VedralMF,CMF}. Even quantum Monte Carlo methods, among the most 
popular techniques known, fails to provide a tool here, due to the intrinsic sign problem commonly arising in such frustrated systems \cite{qmc}. \emph{Tensor network (TN)} techniques, in contrast, are built upon and constructed from notions of entanglement, exhibit neither of the limitations and provide a highly controlled set of powerful 
methods \cite{Orus-AnnPhys-2014,VerstraeteBig,EisertTensorNetworks,AreaReview}. This design feature, thus, renders these techniques particularly  well-suited to study frustrated systems and spin liquids. While such TN algorithms are ubiquitous and widely used in 
systems of one spatial dimension, their higher dimensional counterparts are significantly 
more challenging and complicated in implementation. However, the two-dimensional 
TN states known as \emph{projected entangled pair states (PEPS)} \cite{PEPSOld,iPEPSOld} have recently been maturing 
and can now be reliably used in numerical analysis of two-dimensional systems, 
providing state-of-the-art results when comparing all available methods known to problems \cite{Corboz2DHubbard,CorboztJ}.

In this work, we develop notions of \emph{projected entangled pair states} to new instances of
\emph{projected entangled simplex states (PESS)} \cite{Xiangpess} that are applicable to capture the physics of a quantum magnet Ca$_{10}$Cr$_7$O$_{28}$ \cite{BalzJOP}. This is indeed a quantum material that recently has been studied using several experimental techniques and has been
discovered to have properties of a quantum spin liquid \cite{Balz,BalzPRB}. 
Inelastic neutron scattering revealed that the magnetic excitations are broad at all energies suggesting a multi-particle spectrum consistent with spinons which are the excitations of a quantum spin liquid \cite{Balz}. The spectrum also appeared to be gap-less within the resolution implying that Ca$_{10}$Cr$_7$O$_{28}$ is a gap-less spin liquid. \ak{Earlier theoretical calculations of the dynamical structure factor using functional renormalization group 
methods -- for which unlike here local properties
cannot be computed -- confirms the same property \cite{Balz}.}
This work aims at bringing endeavours of developing tensor network models for strongly correlated systems in two spatial 
dimensions to a new level, giving rise to one of the first instances of studying an actual material that can be prepared in the laboratory using sophisticated tensor-network based schemes -- and not just a paradigmatic model. Thus, our results can be directly compared to experimental data and used as a benchmark in further investigations, providing precisely the type of guidance to experimental efforts 
hinted at above. In what follows, we will start by describing in detail our material Ca$_{10}$Cr$_7$O$_{28}$. We then turn to introducing the numerical technique that we employ to investigate this material. Equipped with this powerful tool, we will demonstrate that it exhibits features of a gap-less quantum spin liquid, a result concomitant with findings in the
experiment, \je{as we explain}. Complementing these results, we will compute the magnetization curve and the magnetic susceptibility of the model in the presence of an external magnetic field, again to good agreement with experiments. \ak{On top of this, we also provide experimental data for the specific heat and the magnetization curve (and susceptibility) which are found to be in good agreement with our numerical calculations and predictions.} In an outlook, we present further perspectives of the approach taken.

\begin{figure}
	\includegraphics[width=0.38\textwidth]{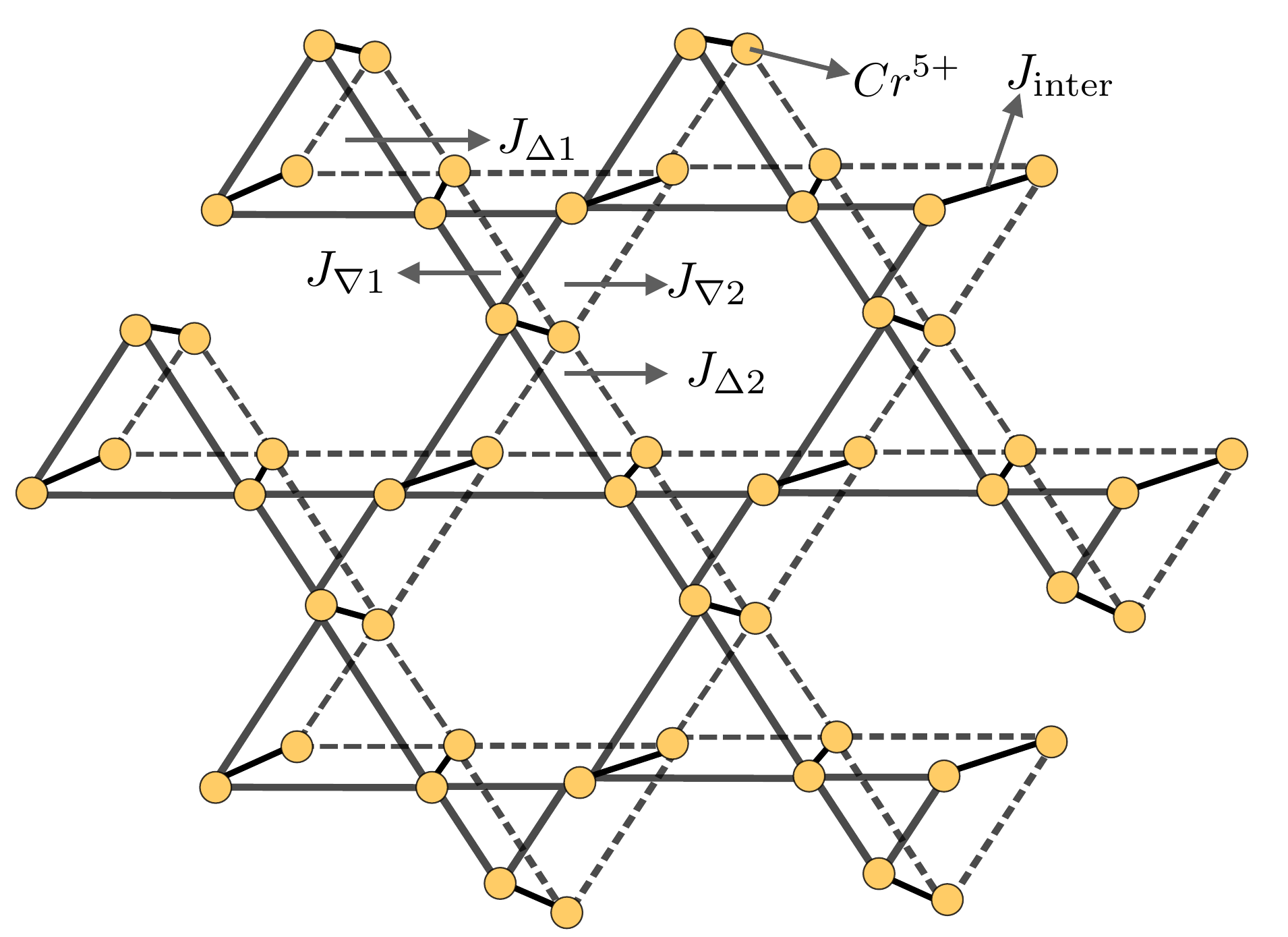}
	\caption{Crystal structure of Ca$_{10}$Cr$_7$O$_{28}$ showing only the magnetic Cr$^{5+}$ ions. It consists of two breathing Kagome layers which are coupled to each other. Each Kagome layer is made up of two inequivalent triangles. The interaction within each triangle is different, both in magnitude and sign, from the rest. The interaction between the spins is of Heisenberg type 
	(Eq.~(\ref{nofield})) and their strengths are given in Table \ref{parameters}.}
	\label{material}
\end{figure}

\emph{Material.} The quantum material that we investigate is known as Ca$_{10}$Cr$_7$O$_{28}$. Powder samples of this material were prepared using solid state synthesis, while single crystals were grown using the travelling solvent floating zone technique. Details on how to prepare this 
compound in the lab can be found in Refs.\ \cite{Balzthesis,BalzJOP}. The crystal structure of this compound has been determined using 
neutron diffraction experiments (\je{all experiments
have been pursued at the Helmholtz Center Berlin}). It has been found to consist of Cr$^{5+}$ ions which have spin-1/2 moments arranged in a 
bilayer breathing Kagome structure \cite{BalzJOP}. The two Kagome layers in this compound are different from each other and each of them again consists of two inequivalent alternating triangles (two different layers of breathing Kagome). The crystal structure of our compound is illustrated in Fig.~\ref{material}.

The Hamiltonian of this compound consists of five inequivalent Heisenberg interactions that can be written as
\beq
\begin{split}
H =  J_{\text{inter}} \sum_{\langle i,j \rangle} \vec{S}_i \cdot \vec{S}_j + J_{\Delta 1}\sum_{\langle i,j \rangle} \vec{S}_i \cdot \vec{S}_j + J_{\nabla 1}\sum_{\langle i,j \rangle} \vec{S}_i \cdot \vec{S}_j \\
 + J_{\Delta 2}\sum_{\langle i,j \rangle} \vec{S}_i \cdot \vec{S}_j
+ J_{\nabla 2} \sum_{\langle i,j \rangle} \vec{S}_i \cdot \vec{S}_j 
\end{split}
\label{nofield}
\eeq
where \ak{$\langle i,j \rangle$ refers to the nearest-neighbor interaction in the lattice.} The first term with interaction strength $J_{\text{inter}}$ corresponds to the coupling between the two layers in the compound and is ferromagnetic in nature. The second term with strength $ J_{\Delta 1}$ corresponds to the coupling of the up triangles in the first layer and is also ferromagnetic in nature. $J_{\nabla 1}$ is for the anti-ferromagnetic down triangle in the first layer. Similarly, for the second layer, the up ($J_{\Delta 2}$) and the down ($J_{\nabla 2}$) triangles have anti ferromagnetic and ferromagnetic coupling, respectively. The nature and the precise strength of these coupling parameters were determined in Ref. \cite{BalzPRB} and are summarized in Tab.\ \ref{parameters}.
\begin{table}[]
	\begin{tabular}{|l|l|l|}
		\hline
		Exchange parameter & Coupling strength (in meV) & Interaction type \\ \hline
		 $J_{\text{inter}}$            & -0.08(4)                   & Ferro            \\ \hline
		$ J_{\Delta 1}$            & -0.27(3)                   & Ferro            \\ \hline
		$J_{\nabla 1}$            & 0.09(2)                    & Anti ferro        \\ \hline
		$J_{\Delta 2}$              & 0.11(3)                    & Anti ferro        \\ \hline
		$J_{\nabla 2}$            & -0.76(5)                   & Ferro         \\ \hline  
	\end{tabular}
\caption{Values and nature of the Heisenberg interaction in the crystal structure of Ca$_{10}$Cr$_7$O$_{28}$ illustrated in Fig.~\ref{material}.}
\label{parameters}
\end{table}
Later on, we will add an external magnetic field along the $S^z$ direction to obtain the magnetization curve of the material as follows
\beq
H_{\text{field}} = H + \sum_i g_s \mu_B h S_i^z
\label{field}
\eeq
where $g_s$ is the g-factor whose value is approximately 2. $\mu_B$ is the Bohr magneton whose value is approximately $5.7883818012 \times 10^{−5}$ eV per Tesla and $h$ is the strength of the external field. We will obtain the ground state of the Hamiltonians in Eq.~\eqref{nofield} and \eqref{field} using our PESS technique.

\emph{Numerical techniques.} 
Both one and two-dimensional TN algorithms have been applied in the past to study paradigmadic two-dimensional frustrated systems \cite{BauerCSL,Xiangkhaf}. PEPS algorithms have also been very successfully used in the past to study the Shastry-Sutherland model which is known to describe the orthogonal-dimer anti-ferromagnet SrCu$_2$(BO$_3$)$_2$ up to a very good approximation where the technique helped to gain new understanding of the magnetization process of the material that were not known before using other tools \cite{SSlandpeps1,SSlandpeps2}, also for the distorted case \cite{SSlandpeps4} as well as in the presence of impurities \cite{SSlandpeps3}.  Two-dimensional 
TN algorithms have also been used to study thermal states \cite{piotr2012,piotr2015,piotr2016,Augustine}, open dissipative systems \cite{Kshetrimayum,Weimeropenreview}, time evolution \cite{piotrevolution,Claudiusevolution,Augustine2DMBL} and even more extensively to study the paradigmatic Kagome Heisenberg anti-ferromagnet,
\cite{Sachdevkhafz2,Gotzekhafz2,dmrgkhafz2,WhiteKagomeqsl,Depenbrockkhafz2,Jiangkhafz2,Rankhafu1,Iqbalkhafu1,Xiangkhaf,Iqbalj1j2u1,KshetrimayumkagoXXZ}.

An intricate feature of the Kagome lattice is the presence of corner sharing triangles. One possibility to study it using 2D Tensor networks is to use conventional pair-wise entangled PEPS directly in the Kagome lattice by either mapping the Kagome lattice onto a square lattice (at the expense of an overhead) or directly updating every two-sites in the Kagome lattice. For instance, mapping the spins in the Kagome to a square lattice as in Refs.\ \cite{KshetrimayumkagoXXZ,Kshetrimayumthesis}, we saw that the local tensors are of the order of $d^3D^8$ and $D^4$. Directly updating two sites without mapping to a square lattice would require local tensors of the order of $D^4d$ and $D^2$. Here $D$ and $d$ correspond to the bond and physical dimension of the Tensor network. On top of this computational bottleneck, neither of these techniques capture the important multi-partite entanglement of the frustrated system in a natural fashion.  The second possibility is to use so-called \emph{projected entangled simplex states (PESS)}\ak{\cite{Xiangpess}}, generalizing PEPS. 
This technique updates three sites in a unit triangle and therefore captures the multi-partite entanglement in a 
very natural fashion. There are advantages in computational effort as well: the local tensors involved are of the order of $dD^2$ and $D^3$. Hence, the latter is more efficient.  It is this
seemingly innocent feature that allows to go to very large bond dimensions. This is also essential if we want to study the entanglement scaling. For these reasons, we will use the 
PESS technique in our study of this double layer Kagome compound. A detailed discussion comparing the two techniques can be found in Ref.\ \cite{KshetrimayumkagoXXZ}. 
Unlike the previous investigations where PEPS/PESS techniques were successfully used to study paradigmatic models like the Kagome Heisenberg anti-ferromagnet \cite{Xiangkhaf,ThibautspinS} or the Shastry-Sutherland model \cite{SSlandpeps1,SSlandpeps2,SSlandpeps3,SSlandpeps4} where the model has later been found to provide very accurate description of the material SrCu$_2$(BO$_3$)$_2$, in our case, our material Ca$_{10}$Cr$_7$O$_{28}$ has first 
been found in the lab to exhibit properties of a quantum spin liquid and its magnetic Hamiltonian has been extracted using inelastic neutron scattering experiments \cite{BalzPRB}. Once the coupling parameters of the Hamiltonian were extracted from the experiments, we investigate it using our PESS approach for these experimental parameters, both in the presence and the absence of external magnetic field. \ak{At the heart of our numerical technique lies the \emph{PESS scheme}
\cite{Xiangpess} which we have adapted for our double layer kagome compound Ca$_{10}$Cr$_7$O$_{28}$} 
 and the \emph{corner transfer matrix renormalization group (CTMRG)} 
\je{\cite{Baxter,Nishino,Nishino2,ctmroman2012,ctmroman2009}
scheme to update \ak{and contract} tensors}
(see the supplemental material for details). \je{In this, we invoke} the simple update to obtain the ground state tensors with six and eighteen site unit cell, \ak{for
reasons that are \je{again} carefully discussed in the supplemental material}. 
\je{At the same time, we} use the CTMRG algorithm to compute the full environment which is then used to calculate the expectation value of the ground state energy, magnetization, etc. Details of the algorithm used and the implementation can be found in the supplementary material. 

\begin{figure}
	\includegraphics[width=0.5\textwidth]{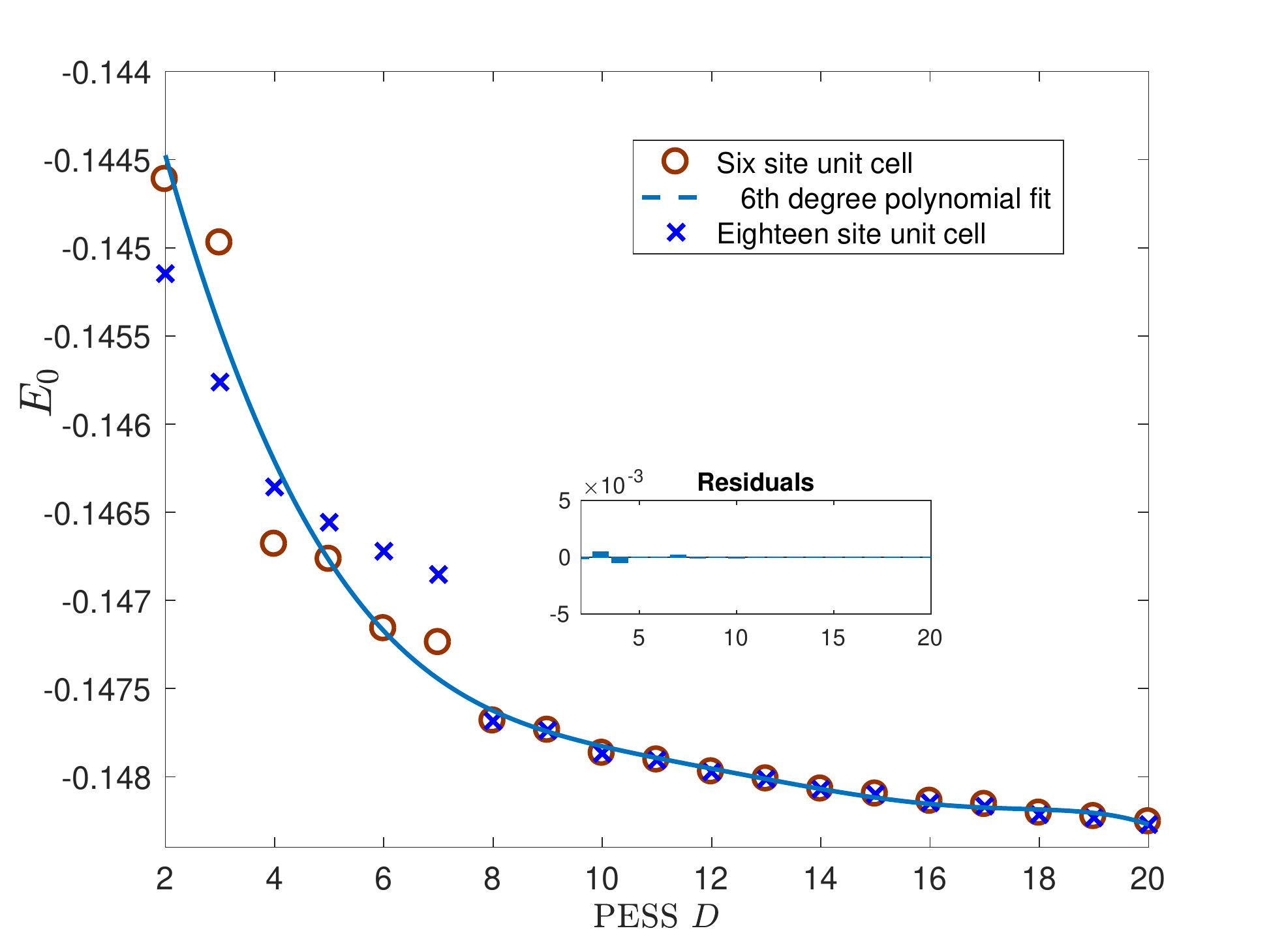}
	\includegraphics[width=0.5\textwidth]{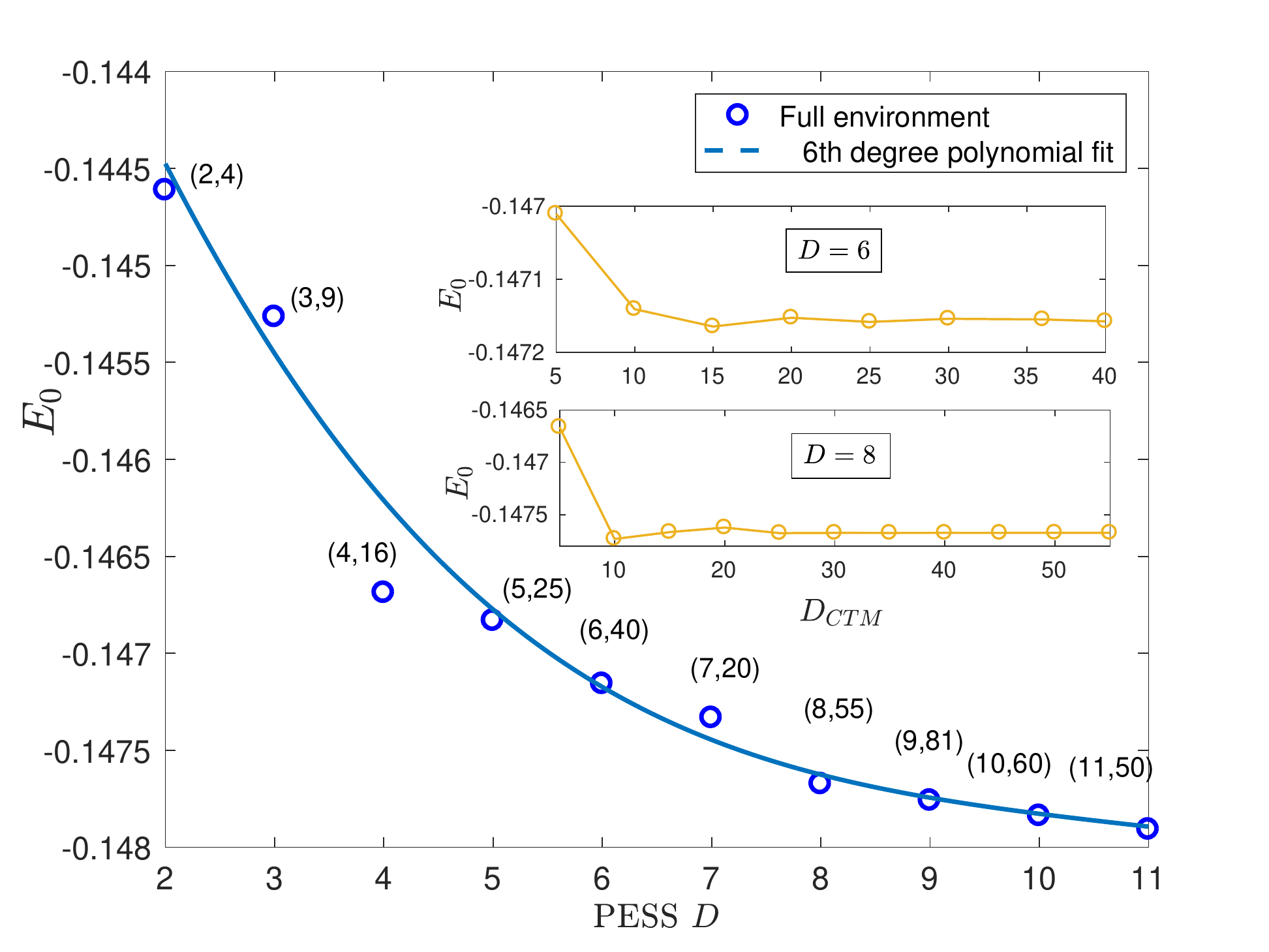}
	\caption{(Top) Ground state energy as function of the bond dimension using the simple environment using a six site unit cell and a eighteen site unit cell. \ak{We also show the fit with a 6th degree polynomial curve (shown in the blue line) along with the errors of the fit.} (Bottom) Same plot using the full environment (CTMRG). The numbers near the data points correspond to the PESS bond dimension $D$ and environment bond dimension $D_{\text{CTM}}$, respectively. (Insets) Ground state energy $E_0$ as function of the bond dimension of the environment $D_\text{CTM}$ for the $D=6$ (top) and $D=8$ (bottom) of the PESS quantum
	state. The results are well converged up to a significant number of digits.}
	\label{simpleenergy}
\end{figure}

\emph{Results for the ground state energy.} To start with, we will compute the ground state energy of our Hamiltonian, i.e., 
${\langle \psi_0 |H|\psi_0 \rangle}/{\langle \psi_0 |\psi_0 \rangle}$ for  $|\psi_0\rangle$ being a good approximation of 
the ground state vector obtained using our update scheme. We will compute this quantity for different values of the bond dimension $D$ of the PESS. We use both the (i) simple environment (SE): that takes into account exact tensor contraction only within a certain cluster
and (ii) full environment (FE): that takes into account the full contraction of the tensors in the thermodynamic limit.  
The results are shown in Fig.~\ref{simpleenergy}.  
From the plots, we can observe an algebraic convergence of the ground state energy as a function of the bond dimension of the PESS $D$ for both the simple environment and the full environment. It is clear that the scaling does not depend on the number of unit cells used in our computation (six and eighteen for SE). Such a behaviour is indicative of the fact that our model is critical or gap-less in nature \cite{algbraicscaling}. For the full environment (bottom), We use the CTMRG algorithm for this purpose with bond dimension $D_{\text{CTM}}$. For each $D$ of the PESS, 
the energies are well converged with the bond dimension of the environment $D_\text{CTM}$. For instance, for $D=6$, the ground state energy is converged up to the 5th decimal place with increasing bond dimension of the environment $D_\text{CTM}$. For $D=8$, the energies are converged up to 6th decimal places with the bond dimension of the environment. This is shown in the insets. The results are similarly well converged up to significant number of digits for other bond dimensions of the PESS $D$. \ak{We have additionally provided a fit using a 6th degree polynomial to better visualize the algebraic convergence of the ground state energy for both the simple as well as the full environment calculations.}

\begin{figure}
	\includegraphics[width=0.44\textwidth]{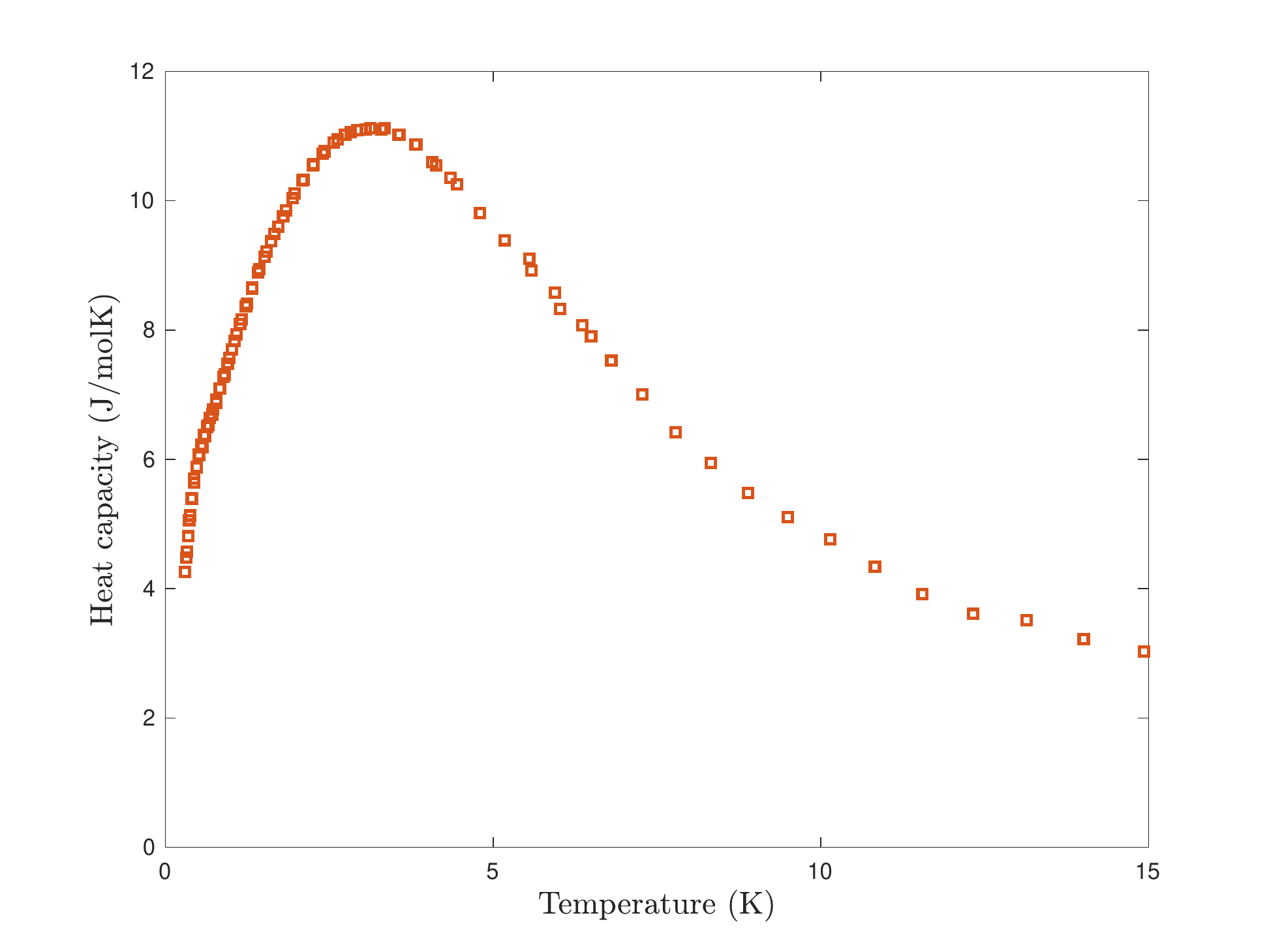}
	\caption{Magnetic heat capacity of Ca$_{10}$Cr$_7$O$_{28}$. A broad maximum is visible at $3$~K while a weak kink appears at $0.4$~K. \ak{Data taken from Ref.~\cite{Balz}}}
	\label{Heatcap}
\end{figure}

\ak{\emph{Results for \je{the} heat capacity.} 
\je{To complement our theoretical findings with experimental results, we}
also show the experimental data for the heat capacity of Ca$_{10}$Cr$_7$O$_{28}$. This is shown in Fig.~\ref{Heatcap}}
\ak{The heat capacity \je{has been} measured on a $0.93$~mg single crystal in the temperature range $0.3-23$~K using a quasi-adiabatic relaxation method in combination with a $^3$He cryostat. A Debye-like phonon contribution $C_p = \alpha T^3$ \je{has been} fitted to the data above $20$~K and subtracted to obtain the magnetic heat capacity. \je{A} magnetization up to $7$~T \je{has been} measured on a $2.05$~mg single crystal using a 
\emph{Quantum Design MPMS 3}
SQUID magnetometer at $1.8$~K.}
%
\cb{\je{The} magnetic heat capacity shown in Fig.\ \ref{Heatcap} reveals no magnetic transitions down to temperatures of $T=300$~mK. A broad peak at $3$~K indicates the onset of short-range correlations and a kink at $0.4$~K suggests a crossover into the fluctuating ground state also evident from muon spectroscopy which revealed the complete absence of any static magnetism even at 19~mK \cite{Balz}. The heat capacity shows no indication of a spin gap which would be observed as a suppression of the heat capacity at lowest temperatures and an exponential increase absent in the data.} \ak{This data 
\je{provides} further
\je{significant evidence for}
the gaplessness of the ground state of the material which we obtained via finite bond dimension scaling of the PESS in the previous section.}

\emph{Results for magnetization and susceptibility.}
We now compute the net magnetization per site of the ground state vector for the parameters in Tab.~\ref{parameters}. This is key to our analysis, as this can be directly measured in the experiment. 
\begin{figure}
	\includegraphics[width=0.45\textwidth]{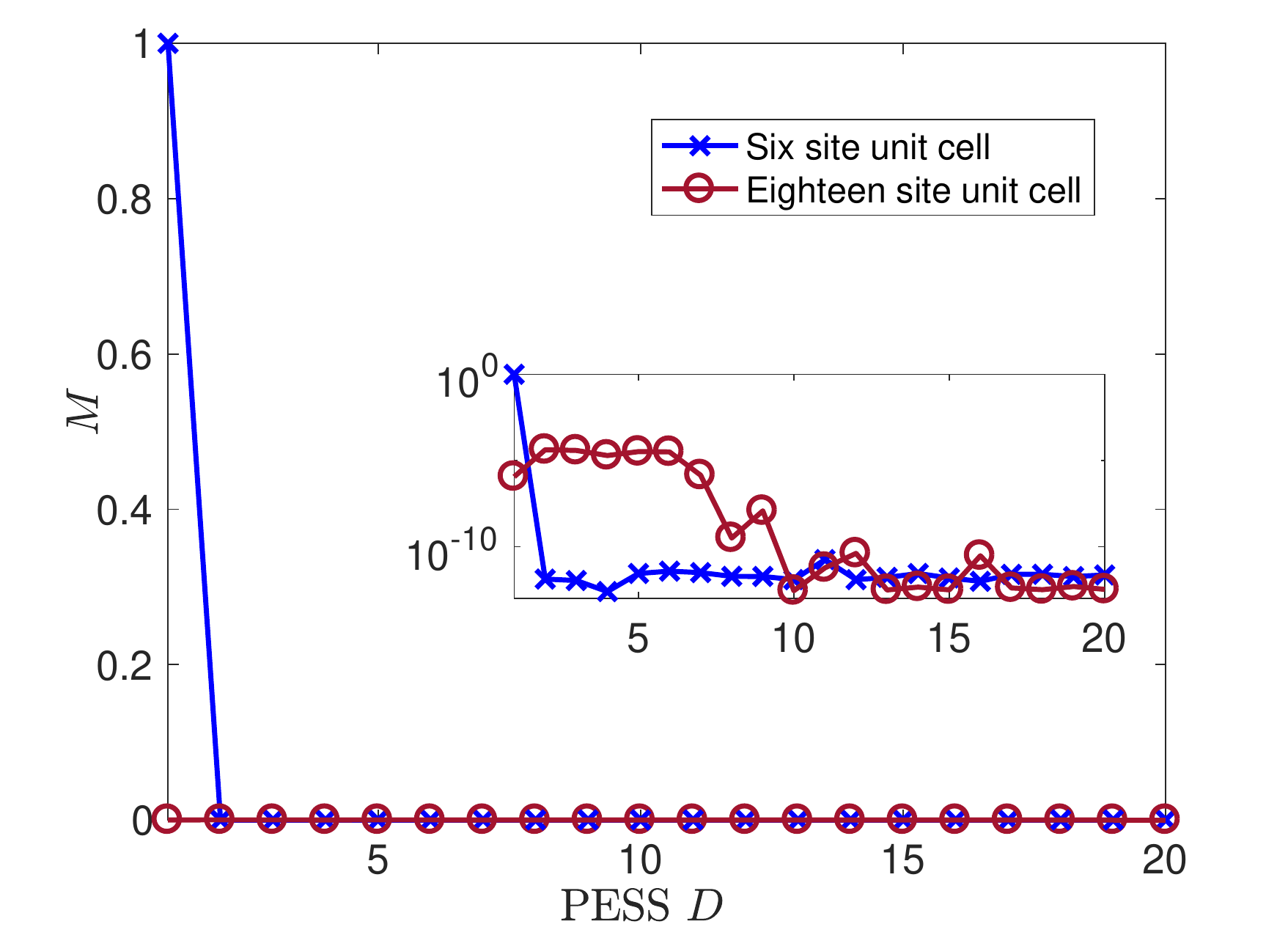}
	\caption{Net magnetization as function of the bond dimension $D$ of the PESS quantum state for the six site and the eighteen site unit cells. Both the plots show the absence of magnetic ordering in any directions. Semi log scales are shown in the insets, respectively.}
	\label{magscaling}
\end{figure}
In Fig.~\ref{magscaling}, we show the average magnetization per site $M = (M_x^2+M_y^2+M_z^2)^{1/2}$ for different 
bond dimensions $D$ of the PESS ground state 
vector, where $M_x, M_y$ and $M_z$ correspond to the expectation value of the Pauli operators. The plots are for six and eighteen site unit cells. 
Again, our results do not depend on the number of unit cells used in our calculations. This is also indicative of a quantum spin liquid preserving the lattice symmetry.
Both calculations have been done using the simple environment. We can see that the net magnetization disappears even for small bond dimensions of the PESS. The absence of symmetry breaking along any directions provides a strong evidence for an $SU(2)$-symmetric quantum spin liquid for the ground state. The algebraic convergence of the energy in Fig.~\ref{simpleenergy} and this absence of magnetic ordering strongly suggests that the ground state is indeed a gap-less spin liquid. \ak{This is further backed up by our experimental data on \je{the heat capacity depicted} in Fig.\ \ref{Heatcap}. The gaplessness of the ground state rules out the possibility of a \emph{valence bond crystal} 
which would have a well defined \je{spectral} gap. Adding more weight to our results, it has been found experimentally that none of the exchange interactions in Ca$_{10}$Cr$_7$O$_{28}$ is able to pair the Cr$^{5+}$ spins into singlets. The triangular motifs prevent the formation of dimers within the Kagome planes, while the interactions that couple the layers into bilayers is ferromagnetic again preventing the formation of dimers. This is the reason that a valence bond solid is suppressed and the spectrum is gapless.} Our \ak{theoretical} results are consistent with experimental findings of the same compound as revealed by several experimental techniques (AC susceptibility, heat  capacity and muon spectroscopy); no static magnetism occurs in Ca$_{10}$Cr$_7$O$_{28}$ even down to temperatures of $T=19$~mK \cite{BalzPRB,Balz}\ak{: all pointing towards a gapless quantum spin liquid as the true ground state of Ca$_{10}$Cr$_7$O$_{28}$ and the main conclusion of this work. A more recent work based on dynamical structure factor and a closer analysis of the heat capacity have come to a similar conclusion about the nature of the ground state of this material \cite{JonasCacro}.}

\begin{figure}
	\includegraphics[width=0.5\textwidth]{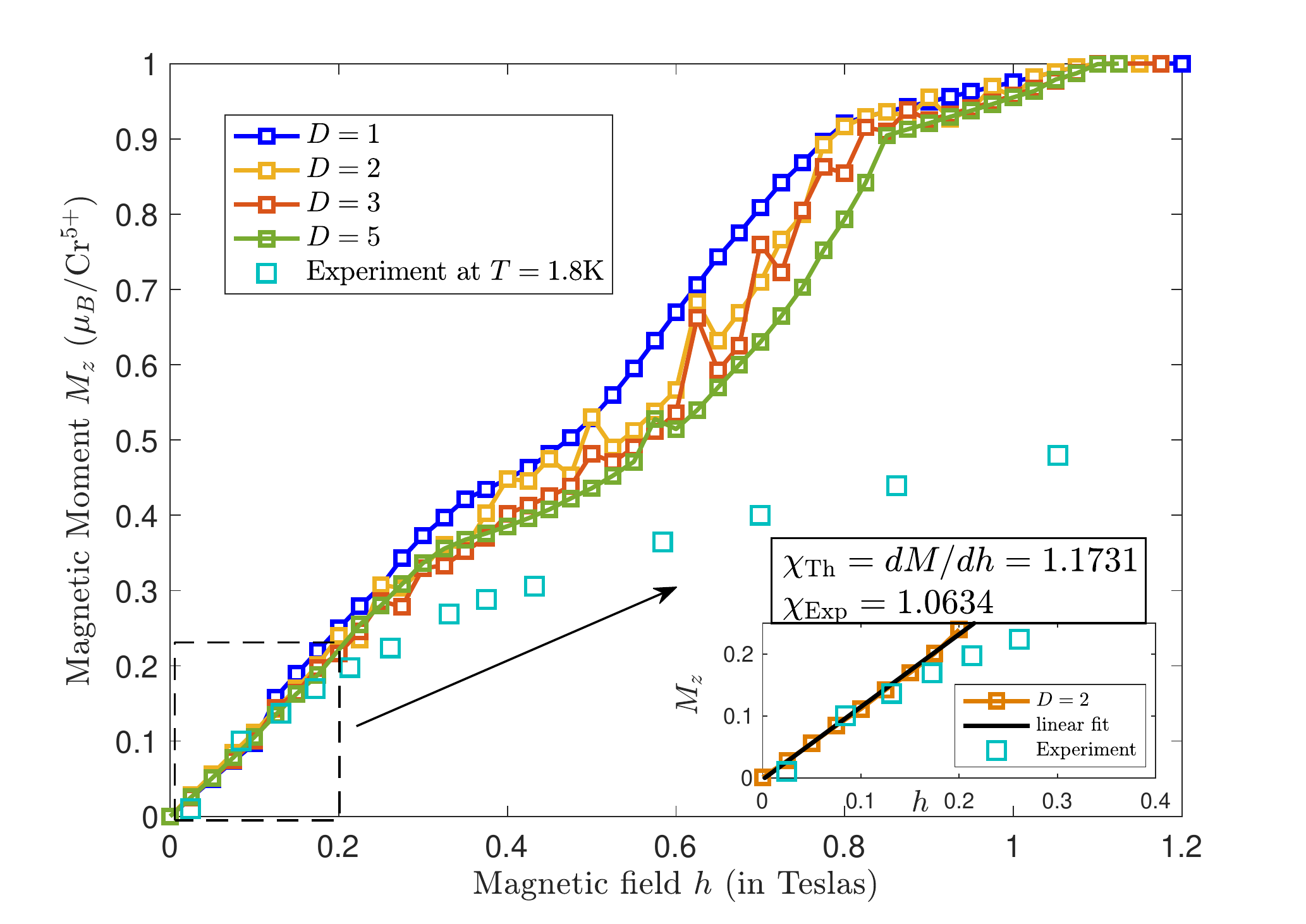}
	\caption{Magnetization curve in the presence of an external field. Slope of the magnetization curve gives us the magnetic susceptibility $\chi$ (inset). \ak{Also shown is the experimental magnetization measured for $H||c$ at $T=1.8$ K.}}
	\label{magnetizationcurve}
\end{figure}

We would now turn to investigating the response of this material to the presence of external magnetic field. To be precise, we would like to find the magnetization curve of the Hamiltonian in Eq.~\eqref{field} in the presence of a strong external field. Our results are \cb{plotted over the experimental data in in Fig.~\ref{magnetizationcurve} \cite{BalzPRB}.}
The plots are obtained using a eighteen site unit cell calculation for $D=1,2,3$ and $5$. We observe a linear curve, without the existence of any magnetization plateaux \ak{unlike} the case of KHAF \ak{\cite{ThibautspinS}}. The linear curve saturates at around 1$T$. The experimental magnetization curve also shows a steep and approximately linear increase in magnetization up to a field of $H=1$T, however, it only achieves ~$50\%$ of the saturation magnetization at this field and continues to increase with a much gentler slope up to saturation at $12$ T (see Ref. \cite{BalzPRB}, Fig.\ 2b). This discrepancy might be due to the fact that the experimentally obtained 
magnetization curve has been measured at $T = 1.8$K and one can expect the behaviour to converge to our theoretically obtained curve as the temperature gets lowered. Our results are also very similar to theoretically obtained mean field results \cite{BalzPRB}. This is to be expected since, in the presence of strong magnetic field, the ground state is in a `product-like' state and \emph{mean field approximations} 
will be very accurate. In fact, the mean field results can be reproduced by our tensor network ansatz for the specific choice 
of $D=1$ (shown by the blue curve in Fig.\ \ref{magnetizationcurve}). In addition to the magnetization curve, we have computed the magnetic susceptibility $\chi$, which is nothing but the slope of the magnetization curve $(dM/dh)$. We obtain a value of $\chi_{\rm Th} = 1.1731$ $\mu_B/(\text{Cr}^{5+}*\text{T})$ \ak{with a fitting error of $8\times 10^{-3}$ at most and convergence \je{up to} four decimal places for the data points} at low values of the magnetic field ($0-0.2$~T). This is \cb{again plotted over the experimental data \ak{$\chi_{\rm Exp}$} in the inset of Fig.~\ref{magnetizationcurve}.
The experimentally obtained value of $\chi_{\rm Exp}$ has been found to be 1.0634~$\pm$ 0.1342 $\mu_B/(\text{Cr}^{5+}*\text{T})$ at $T=1.8$~K which was the lowest temperature measured. It should be noted that the susceptibility was found to increase with decreasing temperature (see Ref.\ \cite{Balz}, Fig.\ 5d), therefore at temperatures below 
$T=1.8$~K larger values of susceptibility are expected. Thus, the higher theoretical value calculated for $T=1.8$~K is in agreement not only on the purely qualitative, but 
also \je{largely} on a quantitative level.}

\emph{Conclusion.}
In this work, we have set out to investigate the properties of the quantum magnet Ca$_{10}$Cr$_7$O$_{28}$
with tensor network methods that are particularly suitable for studying such quantum materials due to the favourable
scaling in effort. We found strong evidence of a gap-less spin liquid from the algebraic convergence of the ground state energy and 
the disappearance of magnetic order parameter using finite entanglement scaling. \ak{Our experimental data on \je{the heat capacity are}  consistent with our theoretical findings.} We have also investigated the 
response of this material to applying an external magnetic field, and have computed the magnetization curve as well as the magnetic susceptibility \ak{both theoretically and experimentally}. Our theoretical results agree well with the available experimental
results including the value of susceptibility measured in experiment at $T=1.8$~K.

Our studies constitutes \ak{one of} the first instances of using state of the art tensor network tools to benchmark the properties \emph{of actual two-dimensional quantum materials} that can be designed and studied 
in the laboratory. In this sense, our study marks a paradigm shift in the treatment of TN tools which were previously used mostly for paradigmatic theoretical models, albeit providing
significant conceptual insight. The highly encouraging results found here provide a road map for future endeavours of 
benchmarking quantum materials with tensor networks. We
plan to make our study even more realistic by studying the finite temperature 
properties of this model so that a one-to-one comparison can be made with the experimental data. Initial steps
taken by us in this direction in developing annealing algorithms for 2D tensor networks \cite{Augustine} suggest
that this can be done. It is the hope that the present work firmly places tensor networks into the portfolio
of experimentalists working with \emph{quantum materials} as well as with \emph{quantum simulators} to benchmark their properties. 

\emph{Acknowledgements.}
A.~K.~would like to acknowledge early discussions with Haijun Liao, Roman Orus and Thibaut Picot for technical details  of the algorithm implemented here. Both J.~E.~and A.~K.~would like to acknowledge discussions with Laura Baez, Alexander Nietner and Emil Bergholtz. We would also like to thank Philippe Corboz for pointing out some of the important references on PEPS. We also acknowledge J{\"o}rg Behrmann and the the HPC Service of ZEDAT, FU Berlin, for providing computing time on the cluster Curta.
This work has been supported by the ERC (TAQ), the Templeton Foundation, and the DFG (CRC 183 Project B1, and EI 519/15-1, EI 519/14-1). This work has also received funding from the European Union's Horizon 2020 research and innovation programme under grant agreement No 817482 (PASQuanS). C.~B.~acknowledges support from the U.S. Department of Energy, Office of Science, Basic Energy Sciences, Division of Scientific User Facilities. This research was also partially supported by the DFG through project B06 of SFB 1143  (project-id 247310070).

\bibliographystyle{apsrev4-1}
%

\section{Supplementary information}
In this supplementary information, we will provide details of our numerical technique and the implementation used.
PESS (PEPS) algorithms involve basically two stages: (i) the update scheme that refines the tensors 
and (ii) the contraction scheme that allow to compute properties. 
The first scheme is when we update the tensors, for example, in this case do an imaginary time evolution to obtain the ground state. The contraction scheme is required because we aim at computing the expectation values once we have the ground state. Let us describe the schemes that we employ in this investigation below.

\subsection{Update scheme}
\begin{figure}
	\includegraphics[width=0.35\textwidth]{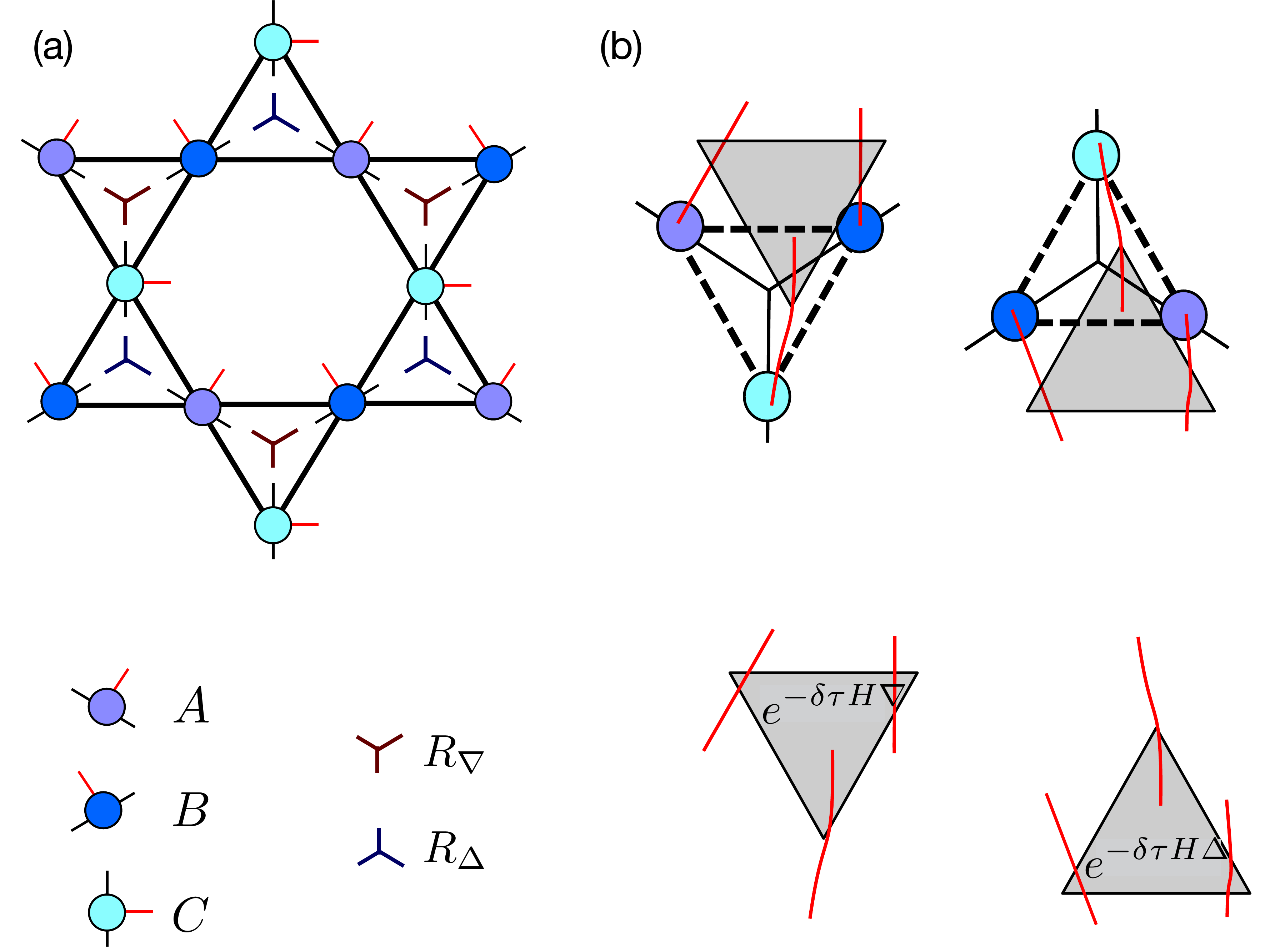}
	\caption{(a) Tensors making up the Kagome lattice in the PESS formalism. (b) Applying an update corresponding to a single
	Trotter step. The grey triangle tensors correspond to the imaginary evolution operator for each cluster.}
	\label{pess}
\end{figure}

In this work, we employ a \emph{simple update scheme} as it is singled out because of its numerical efficiency. \ak{This is of key importance for the problem at hand since we are interested in the scaling behaviour of the ground state energy with large bond dimension. Other optimization schemes such as the full update \cite{iPEPSOld} or the gradient optimization 
as presented in Ref.~\cite{VUMPS2D} cannot be used in this context as they are limited to small bond dimensions that cannot be used to extract the correct scaling behaviour.} However, we will compute the full environment when we calculate all expectation values. In order to make the update, we start by decomposing our Hamiltonian into two parts which do not necessarily commute, i.e., $H = H_{\Delta} + H_{\nabla}$. The first part of the Hamiltonian acts only on the tensors making the up triangle while the second part acts on the down triangle. The algorithm proceeds by applying $e^{-\delta  \tau H_{\Delta}}$ and $e^{-\delta \tau H_{\nabla}}$ successively on some random initial PESS state vector (distributed
according to uniformly distributed random numbers in the interval $(0,1)$). 
This means that the normalized state vector $|\psi \rangle = e^{-\tau H} |\psi_0 \rangle$ in the limit of $\tau \rightarrow \infty$ is expected to provide 
the ground state vector upon convergence, where $|\psi_0 \rangle$ is some initial state vector captured faithfully as a PESS
\cite{Xiangpess}. The same applies for the case of the 9-site (eighteen site) unit cell where we apply the projection operator to six different triangles successively corresponding to nine physical tensors to obtain the ground state vector. 
We use successive Trotter steps of $10^{-1}$, $10^{-2}$, $10^{-3}$ and $10^{-4}$. Thus, the error originating 
from the Trotterization is at most of $O(\delta \tau^2)$ i.e. $O(10^{-8})$.
Thus, once we reach convergence, the ground state vector in the PESS formalism can be 
expressed mathematically for the 3-site 3-PESS as
\begin{equation}
| \psi \rangle = {\rm tr} (X R_{\mu}^{a',b',c'} A^{a',a,i_a} B^{b',b,i_b} C^{c',c,i_c} ) | \dots ,i_a, i_b, i_c ,\dots \rangle
\end{equation}
where $R_{\mu} = R_{\Delta}$ or $R_{\nabla}$, the simplex tensors of dimensions $D\times D \times D$ with $D$, the bond dimension of the PESS, and $X$ denotes the entire rest of the contracted tensor network.
It connects the tensors $A$, $B$ and $C$ associated with physical sites, 
each having physical dimension $d^2$ and bond dimension 
$D$, so that the entire tensors are of degree three and dimension 
$D\times D \times d^2$. It is key to the approach taken here that $d^2$ is the physical dimension of the double layer system of  spin-1/2 systems, here seen as one physical system of local dimension four. This is illustrated in Fig.~\ref{pess}
The above expression is for the 3-site 3-PESS, one can express similarly for the 3-site 9-PESS.

\subsection{Contraction scheme}
Once we have obtained the ground state vector in the form of PESS, we need to be able 
compute expectation values of the Hamiltonian as well as of observables. This can be done by either using a simple 
environment or a full environment of the tensors. We will do both. In the case of the simple environment, 
we neglect correlations beyond a certain cluster and therefore do not need to contract the whole TN in the thermodynamic
limit. \ak{For example, simple environment was used to calculate to ground state of the Heisenberg antiferromagnet on the Huisumi lattice \cite{Xiangkhaf}.}
There are numerical evidences suggesting that such approximations give very accurate results \cite{Thibautnematic}. For the full environment, we need to contract the TN with a bond dimension that is the 
square of the original bond dimension. In both worst case and average case complexity, this is known 
to be a computationally hard problem for exponential precision \cite{Schuchpepshard,Eisertpepshard}. 
This statement can, however, be lifted for ground states of uniformly gapped systems \cite{PhysRevA.95.060102}. 
Indeed, in practice, 
there are several excellent algorithms that can efficiently and accurately approximate the full environment. 
\emph{Corner transfer matrix renormalization group (CTMRG)} methods \cite{ctmroman2012,ctmroman2009}, 
\emph{boundary matrix product states (bMPS)} \cite{iPEPSOld}, and channel environments \cite{VUMPS2D} 
constitute some examples. A detailed work comparing the different contraction schemes will be presented in the future. In this work, we will use the CTMRG technique to compute the environment. We describe it briefly below.

\begin{figure}
	\includegraphics[width=0.48\textwidth]{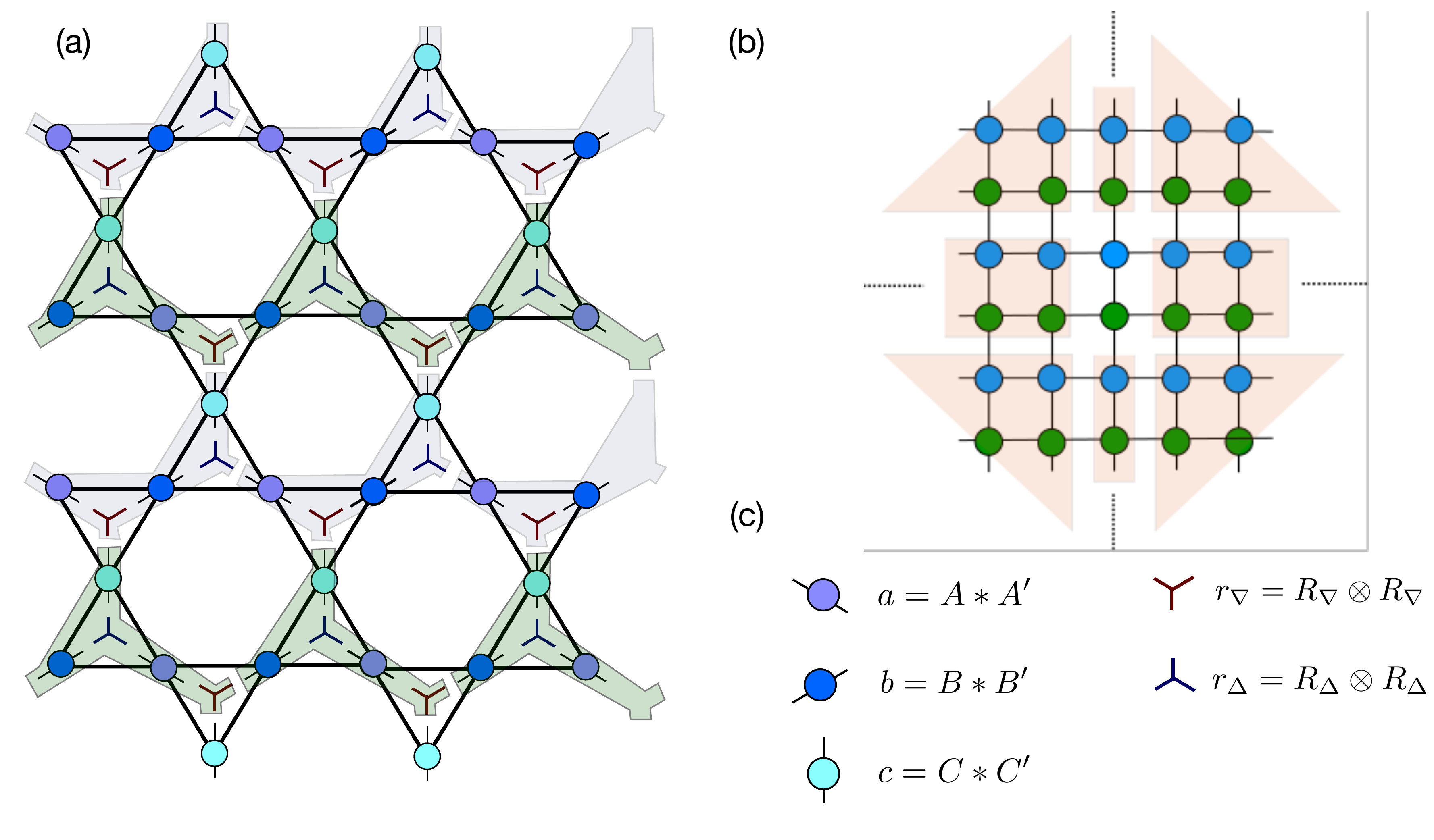}
	\caption{Scheme for PEPS contraction using the full environment. (a) First we do the overlap of the TN corresponding to the state vectors and their duals. Tensors within a shaded region are grouped together. (b) Once, the tensors are grouped as in (a), we obtain a TN for a square lattice. This can be contracted using several techniques such as the CTMRG. The shaded region corresponds to fixed point tensors of the CTM algorithm. (c) Components of the tensors in the Kagome lattice.}
	\label{kagotosquare}
\end{figure}

To start with, we regroup the physical and simplex tensors in the original Kagome lattice, so that we obtain a square lattice which can be contracted in a straight forward manner using the CTMRG technique. Obviously, such a grouping is not unique. We choose the grouping as shown in Fig.~\ref{kagotosquare}.
The TN shown there corresponds to the overlap of the state vector $|\psi \rangle$ and the dual vectors $\langle \psi |$. This means that for a PESS with bond dimension $D$ description of the state vector, we actually need to contract a PEPS with bond dimension $D^2$ to compute the energy or the observable. Fig.~\ref{kagotosquare} shows the 3-site 3-PESS (six site unit cell for our double layer compound). The resulting square lattice from this grouping has a two-site unit cell as shown in Figure \ref{kagotosquare}(b). Once the square lattice is ready, we approximate the tensors in the shaded region in Fig.\ \ref{kagotosquare} using four corner matrices and a number of half-row and half-column transfer matrices depending on the number of unit cells. Details on how to compute these fixed point tensors can be found in Refs.~\cite{ctmroman2012,ctmroman2009}. While computing the environment for a particular site, we fix the bond dimension of the CTM tensors to some number $D_{\text{CTM}}$. This is known as the bond dimension of the environment. The physical dimension of the CTM tensors is given by the bond dimension $D$ of the PESS quantum state. In our case, this value is $D^2$ since we need to compute the overlap of the vectors and dual 
vectors while calculating the expectation values. 

\end{document}